\documentclass[article,10pt]{iopart}
\usepackage[english]{babel}
\usepackage{graphicx}
\usepackage{iopams,cite}

\def\textdegree{$^\circ$}
\def\diameter{$\oslash$}

\begin{document}

\title[]
{The role of rf-scattering in high-energy electron losses from minimum-\textit{B} ECR ion source}
\author{I~V~Izotov$^{1,2}$, A~G~Shalashov$^{1,*}$, V~A~Skalyga$^{1}$, E~D~Gospodchikov$^{1}$, O~Tarvainen$^3$, V~E~Mironov$^{4}$, H~Koivisto$^{5}$, R~Kronholm$^{5}$, V~Toivanen$^{5}$, B~Bhaskar$^{5,6}$}

\address{
 $^1$Institute of Applied Physics of Russian Academy of Sciences, 603950 Nizhny Novgorod, Russia\\
 $^2$Lobachevsky State University of Nizhny Novgorod, 603950 Nizhny Novgorod, Russia\\
 $^3$STFC, ISIS Pulsed Spallation Neutron and Muon Facility, Rutherford Appleton Laboratory,
 Harwell OX11 0QX, United Kingdom\\
 $^4$Joint Institute for Nuclear Research, 141980 Dubna, Russia\\
 $^5$University of Jyväskylä, 40500 Jyväskylä, Finland\\
 $^6$Univ. Grenoble Alpes, CNRS, Grenoble INP, LPSC-IN2P3, 38000 Grenoble, France
}%

\ead{$^*$ags@ipfran.ru}
\vspace{10pt}

\begin{indented}
\item[]{\today}
\end{indented}

\begin{abstract}

The measurement of the axially lost electron energy distribution escaping from a minimum-\textit{B} electron cyclotron resonance ion source in the range of 4--800 keV is reported. The experiments have revealed the existence of a hump at 150--300 keV energy, containing up to 15\% of the lost electrons and carrying up to 30\% of the measured energy losses. The mean energy of the hump is independent of the microwave power, frequency and neutral gas pressure but increases with the magnetic field strength, most importantly with the value of the minimum-\textit{B} field. Experiments in pulsed operation mode have indicated the presence of the hump only when microwave power is applied, confirming that the origin of the hump is rf-induced momentum space diffusion.
Possible mechanism of the hump formation is considered basing on the quasi-linear model of plasma-wave interaction.

\end{abstract}

\submitto{\PPCF}

\maketitle
\ioptwocol

\section{Introduction}

Electron cyclotron resonance (ECR) ion sources are widely used as injectors of multicharged ions into accelerators -- many of remarkable results in fundamental nuclear physics research derive from significant progress of the ECR ion sources over the past several decades. Improvements of the magnetic plasma confinement have yielded significant enhancements of the ECR ion source performance. However, despite the fact that the enhancement of the magnetic systems leads to an obvious increase of the ion confinement time, production of high charge state ions is impossible if the mean electron energy is too low for its efficient stripping. Thus, knowledge on electron energy distribution is essential for evaluation the ionization rate of a particular charge state and, finally, tuning the ion source for the best performance. Furthermore, it has been shown that in  ECR ion source, the  plasma is prone to cyclotron kinetic instabilities, which are widely recognized as a factor limiting the ECR ion source performance, i.e.\ in the high charge state current \cite{PSST2014, RSI2015}. The onset of kinetic instabilities is determined by the full electron distribution function of confined electrons in 6D phase-space, which highlights the necessity of gathering all possible information, including the electron energy distribution, to understand the underlying mechanism and to mitigate the instabilities.

A traditional technique of estimating the electron energy distribution in ECR ion source is based on measurement of the plasma and wall bremsstrahlung. However, such measurements usually give rather crude information, a spectral temperature and maximum electron energy. To complement this technique, direct measurements of the lost electrons energy distribution (LEED) have been introduced resolving the energy distribution of electrons escaping axially from an open magnetic trap \cite{PSST2018}; an overview of the recent experiments may be found in \cite{Izotov2020}. It is emphasized that the distribution function of confined electrons in a magnetic trap and the LEED are very likely different; however, it may be argued that the LEED reflects sensitively the distribution of the confined electrons, and thus may be used for validation of physical models covering both confined and lost particles.

The present paper elaborates on previous research on the electron energy distribution in high-frequency minimum-\textit{B} ECR ion source reported in \cite{PSST2018}. The measurements reported hereafter, being more precise and extensive, confirm the characteristics of the energy distribution of electrons escaping the magnetic trap obtained previously both qualitatively and quantitatively. Furthermore, the results reveal hitherto unexplored features of the energy distribution that allow extending the knowledge of basic plasma physics of ECR-heated plasmas.

Compared to \cite{PSST2018},  the same apparatus was used in the present study, though the measurement technique was noticeably enhanced, resulting in a wider energy range available for the analysis, being now from 4 keV up to 450-800 keV depending on the dataset. This was done to focus on the high energy tail of the lost electron energy distribution, as the earlier measurements, limited to energies below 250~keV, had suggested a hump at energies exceeding 200~keV \cite{PSST2018}. Although it could be argued that the electrons with energies on the order of hundreds of keV play a minor role in ionization due to small electron impact ionization cross-section at relativistic energies, their energy distribution is still of great interest. In particular, high-energy electrons are considered to be responsible for the onset of cyclotron instabilities in ECR ion sources \cite{Shalashov-REV}. Accumulation and  losses of such energetic electrons are relevant for the local structure of the plasma potential and are believed to influence the overall plasma confinement including electrostatic trapping of the high charge state ions \cite{Pastukhov}. Therefore, investigation of a high-energy tail of the LEED is considered to be of fundamental interest.

\begin{figure*}[tb]
\includegraphics[width=1.0\linewidth]{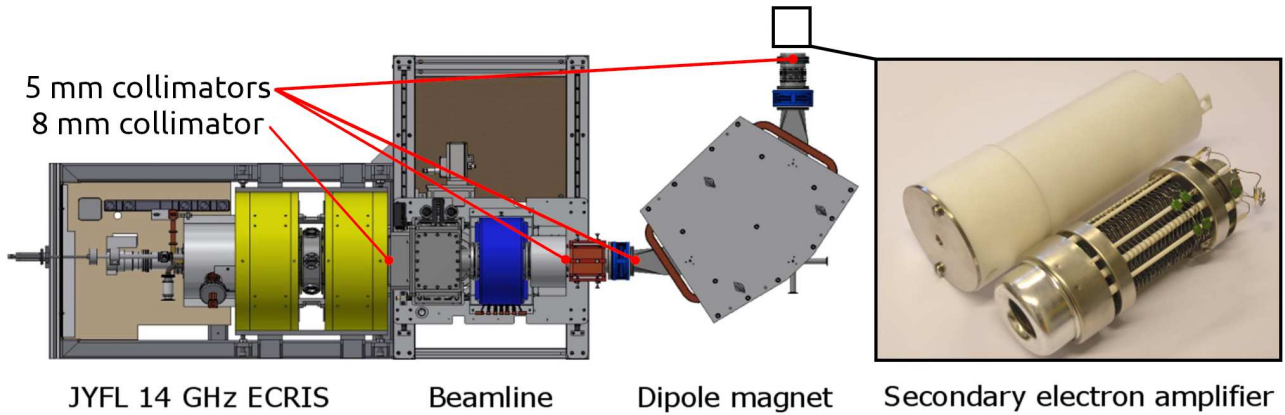}
\caption{\label{fig1} A schematic view of the experimental setup. From left to right: the JYFL ECR ion source, low energy beamline with 5 mm collimators placed between the solenoid (blue) and dipole magnets, the 90\textdegree ~dipole magnet used as an electron spectrometer and the secondary electron amplifier placed at the end of the displayed beamline section following the dipole magnet. The insulating cover (white) and \diameter=5 mm entrance collimator of the amplifier have been removed for illustration purposes to expose the amplifier chain. The position of the collimators is marked with red lines.}
\end{figure*}

\section{Experimental setup}
The experiments were performed with the 11-14 GHz A-ECR-U type ion source at the Accelerator Lab of the University
of Jyv¨askyl¨a ion source (JYFL ECR ion source)  \cite{JYFL_ECRIS}, shown schematically in \fref{fig1}. The source uses an Nd-Fe-B permanent magnet sextupole arrangement and two solenoid coils to form a minimum-\textit{B} structure for plasma confinement. The axial field strength can be varied by adjusting the solenoid currents, which affects the injection ($B_\mathrm{inj}$) and extraction ($B_\mathrm{ext}$) field values, as well as $B_\mathrm{min}/B_\mathrm{ECR}$  (with $B_\mathrm{ECR}=0.5$ T at 14 GHz). In other words, $B_\mathrm{min}$ cannot be adjusted independently unlike in most high-performance ECR ion sources equipped with more than two coils. The reader is referred to \cite{PSST2018} for more details on the magnetic field configuration. The solenoids were adjusted both simultaneously and independently, which allowed a comparison of the magnetic field configurations with (almost) the same $B_\mathrm{min}$ but having different $B_\mathrm{inj}$ and $B_\mathrm{ext}$, thus testing certain hypotheses about physical processes affecting the lost electron energy distribution. The electrons in the ECR plasma were heated either by 10--600 W of microwave power at 14 GHz provided by a klystron amplifier, or 100--200 W of microwave power in the range of 11.4--12.7 GHz using a TWT amplifier connected to a secondary waveguide port.

Typical neutral gas pressures were in the $10^{-7}$~mbar range. Oxygen was used in the present study to make it coherent with the previous ones, though some data were acquired with argon to confirm the observations and test for their dependence on a gas composition. The electron flux escaping the confinement through the circular extraction aperture (\diameter=8 mm) was gathered with two \diameter=5 mm collimators placed between the ion source and the 90\textdegree ~bending magnet used as an electron spectrometer. The electrons were finally detected with a secondary electron amplifier equipped with yet another \diameter=5 mm entrance collimator, placed in the beamline downstream from the bending magnet. The extraction electrodes normally used for optimizing the ion beam optics and the above aluminium collimators were all grounded and all magnetic components (solenoids and $XY$-magnets) were turned off to avoid steering or focusing  the electrons and thereby affecting their energy-dependent transmission probability from the ion source to the detector.

The polarity of the bending magnet power supply was changed from the normal operation where the magnet is used for $m/q$-separation of high charge state positive ions. The magnetic field deflecting the electrons was measured with a calibrated Hall probe. The energy distribution of the electrons precipitating from the trap was then determined by ramping the field of the bending magnet, detecting the electron current from the amplifier with a picoampermeter and applying a set of corrections, taking into account the transport efficiency, electron backscattering and secondary electron yield. Further details of the data processing can be found from \cite{PSST2018}.

\section{Experimental results}
\subsection{Experiments with continuous microwave injection}

The LEED was measured as a function of the microwave power, microwave frequency, magnetic field strength and gas pressure in the stable operating regime (unless otherwise stated) of oxygen and argon plasma.

\begin{figure}
\includegraphics[width=\linewidth]{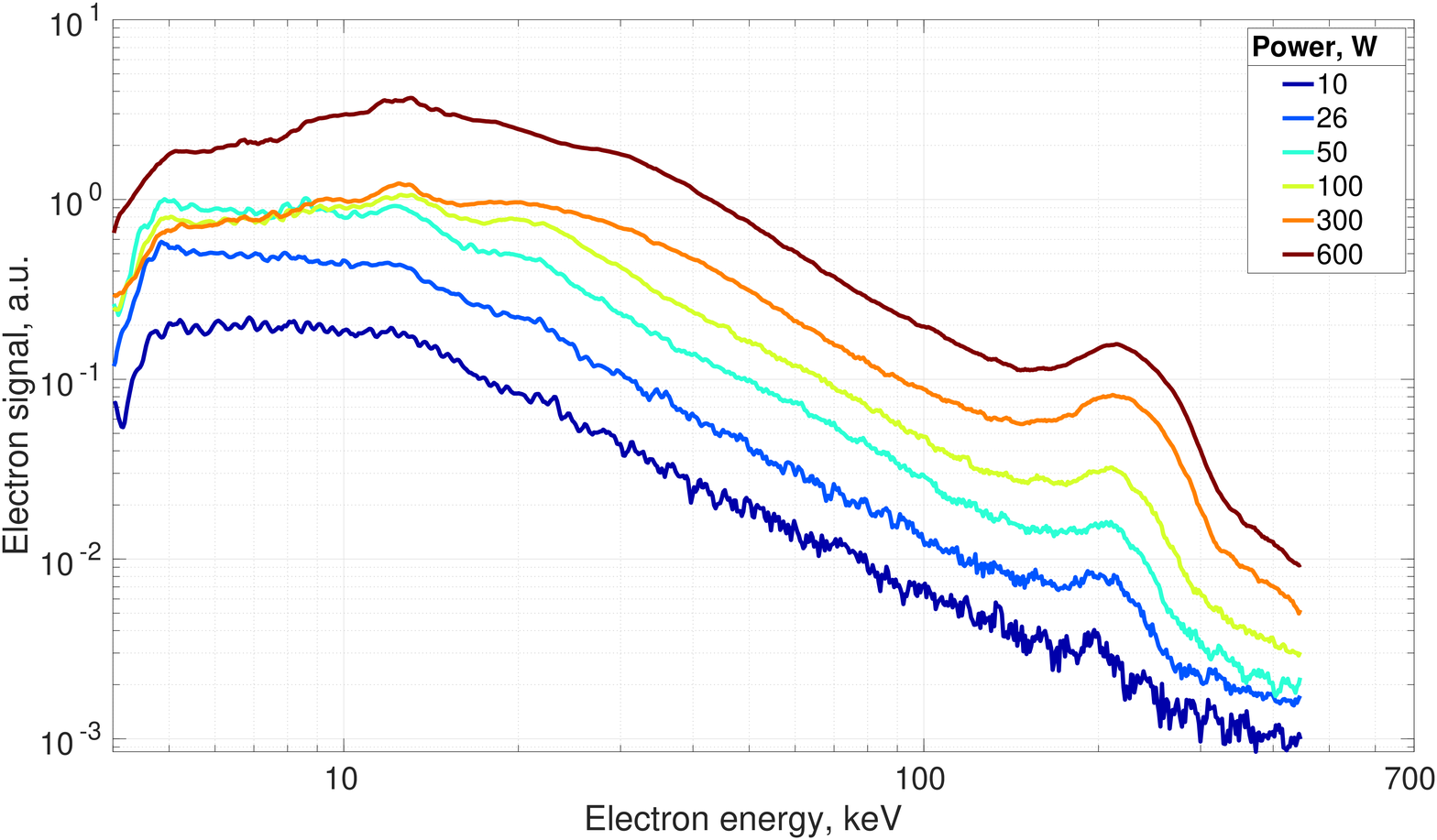}
\caption{\label{fig2}
The electron current (arbitrary units) as a function of the electron energy at different microwave powers. $B_\mathrm{inj}=1.979$ T, $B_\mathrm{min}=0.376$ T, $B_\mathrm{ext}=0.916$ T, $B_\mathrm{min}/B_\mathrm{ECR}=0.753$, oxygen pressure $3.4 \cdot 10^{-7}$ mbar.
}
\end{figure}

\begin{figure}
\includegraphics[width=\linewidth]{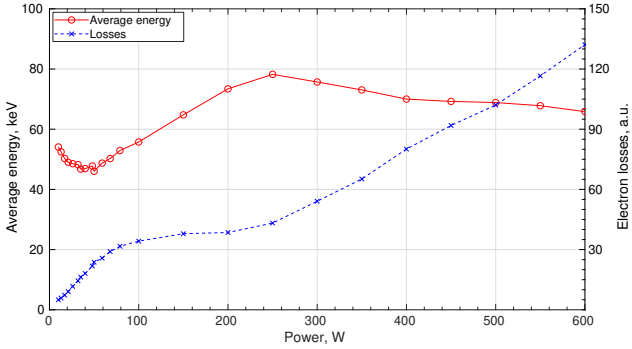}
\caption{\label{fig3}
Total electron losses $F=\int_{\varepsilon_{\min}}^{\varepsilon_{\max}} f(\varepsilon) \:d\varepsilon$ (dashed blue) and the average energy of the escaping electrons $\varepsilon_\mathrm{avg}=F^{-1} \int_{\varepsilon_{\min}}^{\varepsilon_{\max}} f(\varepsilon) \:\varepsilon d\varepsilon$ (solid red) as a function of microwave power calculated for the data in \fref{fig2}. Here  $\varepsilon$ is the electron energy, $f(\varepsilon)$ is the measured LEED, the energy range is defined by $\varepsilon_{\min}=4$ keV  and $\varepsilon_{\max}=800$ keV.
}
\end{figure}

\Fref{fig2} shows the electron current (arbitrary units) as a function of the electron energy at different microwave powers in the range of 10 - 600 W at 14 GHz. Both coils were operated with an equal current of 510 A, yielding $B_\mathrm{inj}=1.979$ T, $B_\mathrm{min}=0.376$ T, $B_\mathrm{ext}=0.916$ T and  $B_\mathrm{min}/B_\mathrm{ECR}=0.753$. The injection field is stronger in spite of the equal coil current due to the presence of an iron plug shaping the field at the injection. The oxygen pressure (hereafter measured without plasma) was
$3.4 \cdot 10^{-7}$ mbar. Total electron losses and the corresponding average energy of the escaping electrons are plotted in \fref{fig3}. Hereinafter, total losses are understood as losses in the energy range 4--800 keV available for measurement.

The most pronounced feature in LEED is the magnitude of a high-energy hump observed at energy of $\sim$200 keV. The magnitude of the hump increases noticeably with power, whereas the overall LEED shape is preserved. The hump contains 10-15\% of the total electron flux and accounts for more than 30\% of measured energy losses, which makes it of fundamental interest as discussed in the following sections. Another clearly distinguishable change occurs when the microwave power changes from $50$ to $250$ W. While the magnitude of the LEED in the range $\varepsilon>30$ keV grows monotonically with power, it exhibits a drop in the range $\varepsilon<30$ keV, which leads to a growth in the average energy seen in \fref{fig3}, whereas the total losses are almost constant in this power range. Both at low ($<$50 W) and high ($>$250 W) power, the average energy decreases while the total losses increase with power, the latter being expressed by the whole curve shifting up in \fref{fig2}. It is unclear from the data in \fref{fig3} whether the growth of the total losses and the hump itself are associated with an increase in plasma density (thus, enhancing collisional losses), which apparently follows from the increase of the power, or due to an increase in rf-induced loss rate, which also grows with the magnitude of the microwave field \cite{RF-1,RF-2,SA-tokman,RF-3}.

Both, the total losses and the average energy, exhibit a change of behaviour at power of $\sim$250 W: the total losses reach a local plateau between 100 and 250 W and then continue to grow monotonically at higher power; meanwhile the average energy starts to decrease slowly when the power exceeds 250 W. The saturation with the microwave power has been associated elsewhere with the saturation of the plasma energy content \cite{Noland-PSST} and confirmed with the JYFL ECR ion source by measuring the volumetric rate of inner shell ionization (through the detection of characteristic x-ray emission) with the ionization rate per unit of absorbed power saturating at 250 W \cite{Sakildien-NIMA} independent of the neutral gas pressure. Altogether these experimental findings and the observed decrease or increase of the LEED average energy and electron flux imply that the microwave power becomes insufficient to maintain a certain EED at the saturation point.

\begin{figure}
\includegraphics[width=\linewidth]{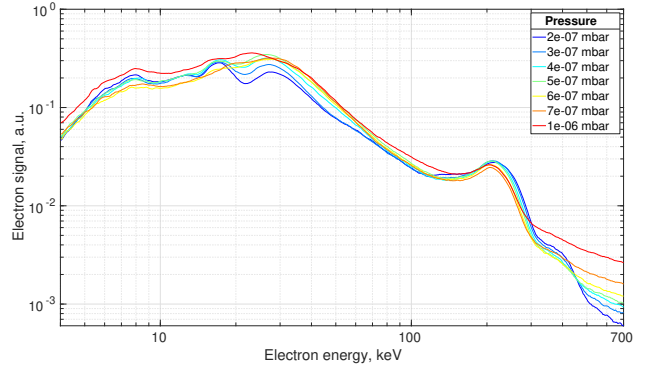}
\caption{\label{fig4}
The electron current (arbitrary units) as a function of the electron energy at different oxygen pressures. $B_\mathrm{inj}=1.960$ T, $B_{\min}=0.367$ T, $B_\mathrm{ext}=0.900$ T, $B_\mathrm{min}/B_\mathrm{ECR}=0.735$, microwave power 300 W at 14 GHz.
}
\end{figure}

\begin{figure}
\includegraphics[width=\linewidth]{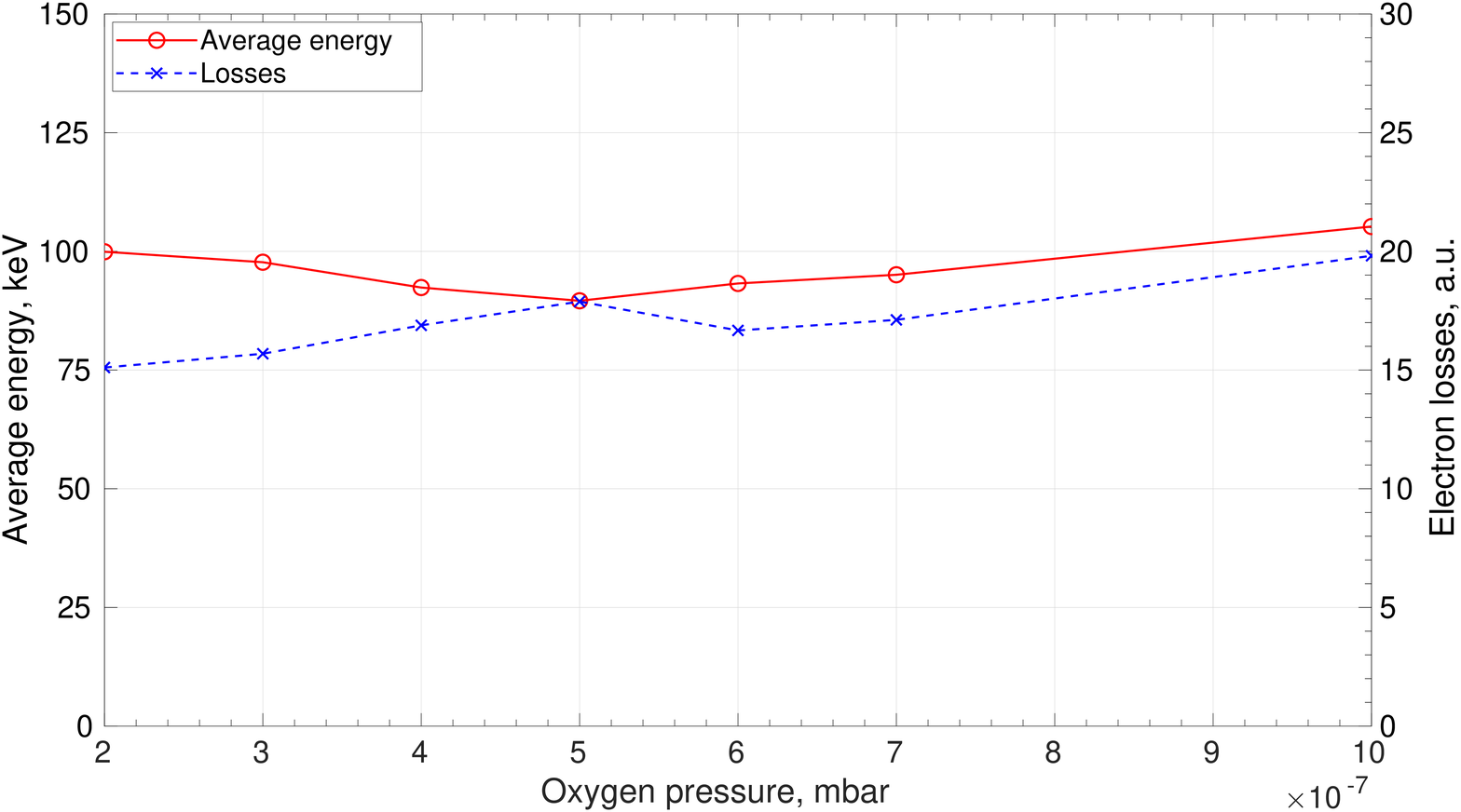}
\caption{\label{fig5}
Total electron losses (dashed blue) and the average energy of the escaping electrons (solid red) as a function of oxygen pressure calculated for the data in \fref{fig4}.
}
\end{figure}

The dependence of LEED on the oxygen pressure is shown in \fref{fig4} with the corresponding average energy and total losses plotted in \fref{fig5}. Neither the LEED nor the average energy and total losses  in the energy range of 4--800 keV change noticeably with the gas pressure. The gas pressure apparently affects the plasma density and, therefore, the observation suggests that the plasma density hardly affects the electron losses in the energy range probed here.

\begin{figure}
\includegraphics[width=\linewidth]{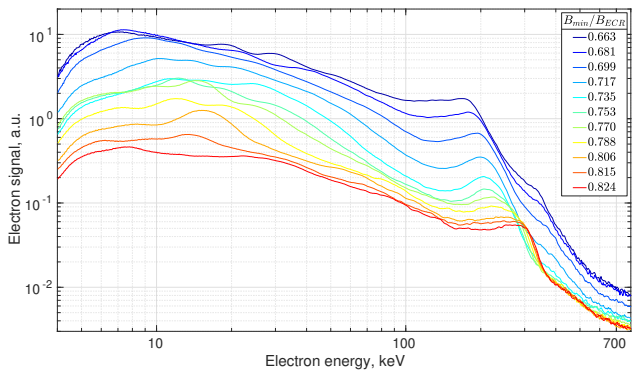}
\caption{\label{fig6}
The electron current (arbitrary units) as a function of the electron energy at different $B_\mathrm{min}/B_\mathrm{ECR}$, microwave power 400 W at 14 GHz, oxygen pressure $3.5 \cdot 10^{-7}$ mbar.
}
\end{figure}

\begin{figure}
\includegraphics[width=\linewidth]{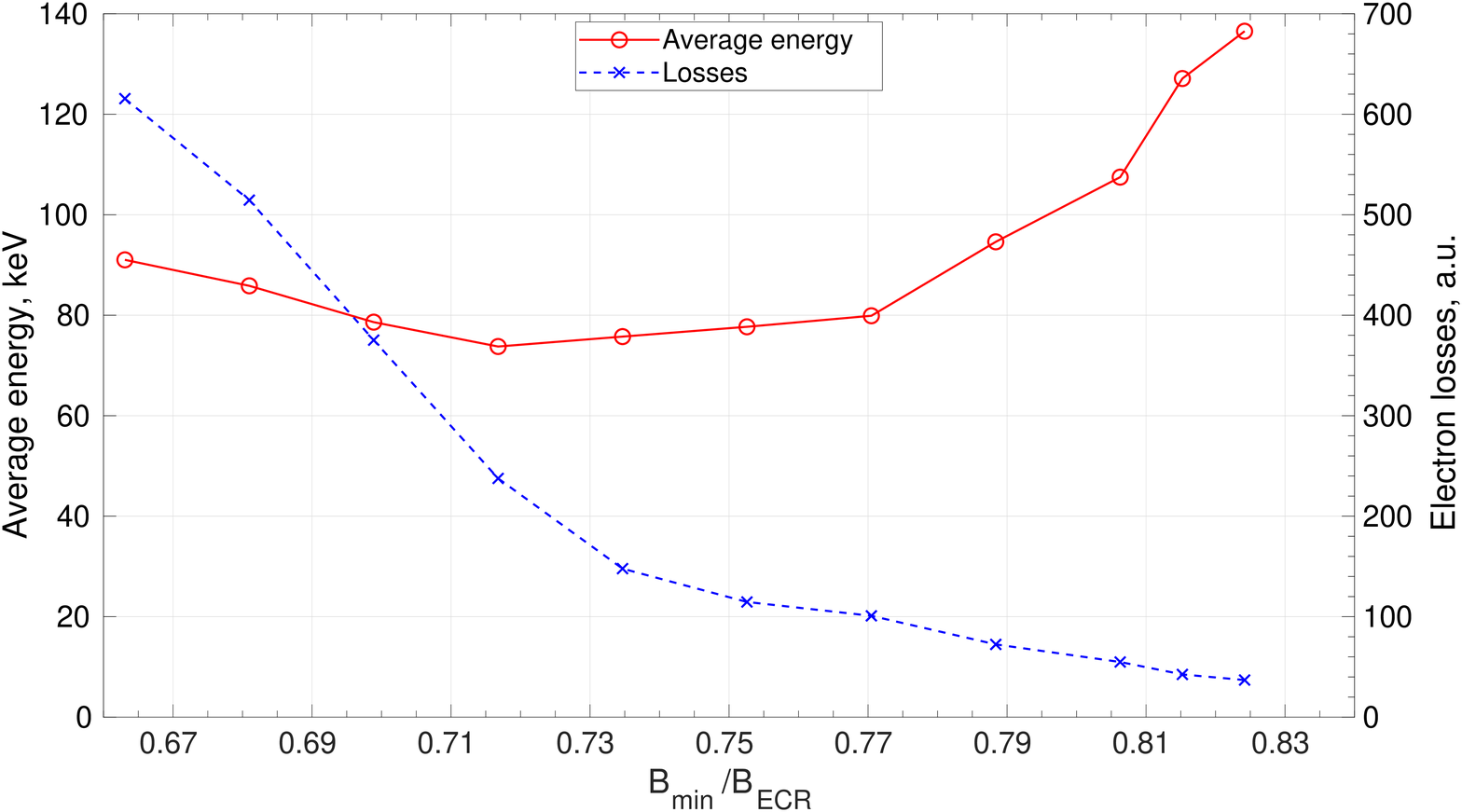}
\caption{\label{fig7}
Total electron losses (dashed blue) and the average energy of the escaping electrons (solid red) as a function of $B_\mathrm{min}/B_\mathrm{ECR}$ calculated for the data in \fref{fig6}.
}
\end{figure}

The dependence of LEED on the magnetic field strength, expressed in terms of $B_\mathrm{min}/B_\mathrm{ECR}$, in the energy range of 4--800 keV  is shown in \fref{fig6}. Both, the average energy and total losses, depicted in \fref{fig7}, clearly depend on the magnetic field. The magnetic field here was changed by adjusting the current in both coils symmetrically, yielding $B_\mathrm{inj}=1.88$ T, $B_\mathrm{min}=0.33$ T, $B_\mathrm{ext}=0.84$ T for $B_\mathrm{min}/B_\mathrm{ECR}=0.663$ and $B_\mathrm{inj}=2.05$ T, $B_\mathrm{min}=0.41$ T, $B_\mathrm{ext}=0.97$ T for $B_\mathrm{min}/B_\mathrm{ECR}=0.824$ at the extremes of the range used here. The increase of the magnetic field by 15\% decreases the axial electron losses (with energies above 4 keV) by more than an order of magnitude and causes the average energy of those electrons to increase by a factor of 1.5. The increase of the magnetic field strength at constant resonant field (microwave frequency) obviously lowers the magnetic field gradient at the resonance zone, which presumably leads to a more effective ECR interaction \cite{Lichtenberg-1991} and an increase in the average electron energy. On the other hand, the increase of the magnetic field leads to more electrons being trapped. Experimental observations are in good agreement with the above reasoning.

Contrary to microwave power and neutral gas pressure having a negligible effect on the energy of the LEED hump, its position shifts considerably with magnetic field; from $\sim$170 keV at $B_\mathrm{min}/B_\mathrm{ECR}=0.663$ to $\sim300$ keV at $B_\mathrm{min}/B_\mathrm{ECR}=0.824$.

The data shown in the present article as well as in the previous study \cite{PSST2018} implies that the magnetic field is the most influential ECR ion source parameter affecting the LEED. This is best visible at high energies including the hump of the distribution. It is worth noting that the high-energy hump has been also observed in wall-bremsstrahlung spectra of the same ion source \cite{XRAY-1, XRAY-2} at similar energies of $\sim200$ keV, shifting towards higher energies with the increase of magnetic field but remained unaffected by the change of microwave power or neutral gas pressure. Although the bremsstrahlung spectrum does not yield unambiguous information on the electron energy distribution and is affected by the collimation, the presence of the hump in the wall bremsstrahlung spectrum allows one to expect the presence of a similar hump in the energy distribution of the lost electrons causing the bremsstrahlung, which is exactly the observation reported here.

In order to determine which magnetic field-related quantity is responsible for the hump energy, a series of experiments was conducted measuring the LEED with different combinations of magnetic field and microwave frequency/power. The magnetic field was varied by tuning the coils independently. In this case the change of the extraction coil current affects the $B_\mathrm{min}$ and $B_\mathrm{ext}$ keeping $B_\mathrm{inj}$ almost unaffected, whereas the change of the injection coil current keeps $B_\mathrm{ext}$ roughly constant, but affects $B_\mathrm{min}$ and $B_\mathrm{inj}$. Operating the coils at different combinations of currents therefore allows one to identify the most influential magnetic field parameter on the hump of the LEED. The heating frequency was varied by using either the 14 GHz klystron or the TWT amplifier tuned to 11.4 GHz, 11.56 GHz, 11.7 GHz, 12.3 GHz or 12.7 GHz at different power levels. Only single frequency heating mode was investigated and the TWT frequencies were selected by minimizing the reflected power. Oxygen with $3.5 \cdot 10^{-7}$ mbar pressure was used in these experiments. Finally, argon with the same pressure of $3.5 \cdot 10^{-7}$ was used at 14 GHz.

\begin{figure}
\includegraphics[width=\linewidth]{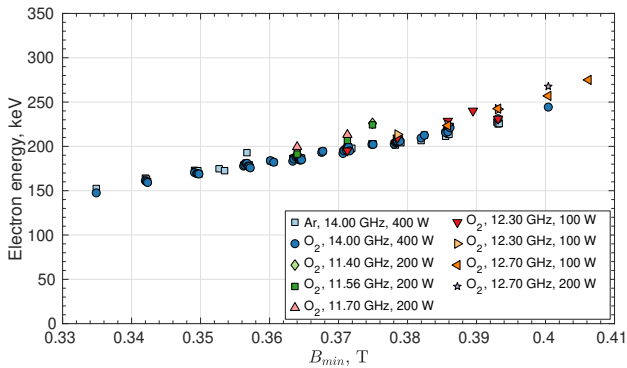}
\caption{\label{fig8}
The electron energy at the peak of the LEED hump as a function of $B_\mathrm{min}$ at
different settings.
}
\end{figure}

\begin{figure}
\includegraphics[width=\linewidth]{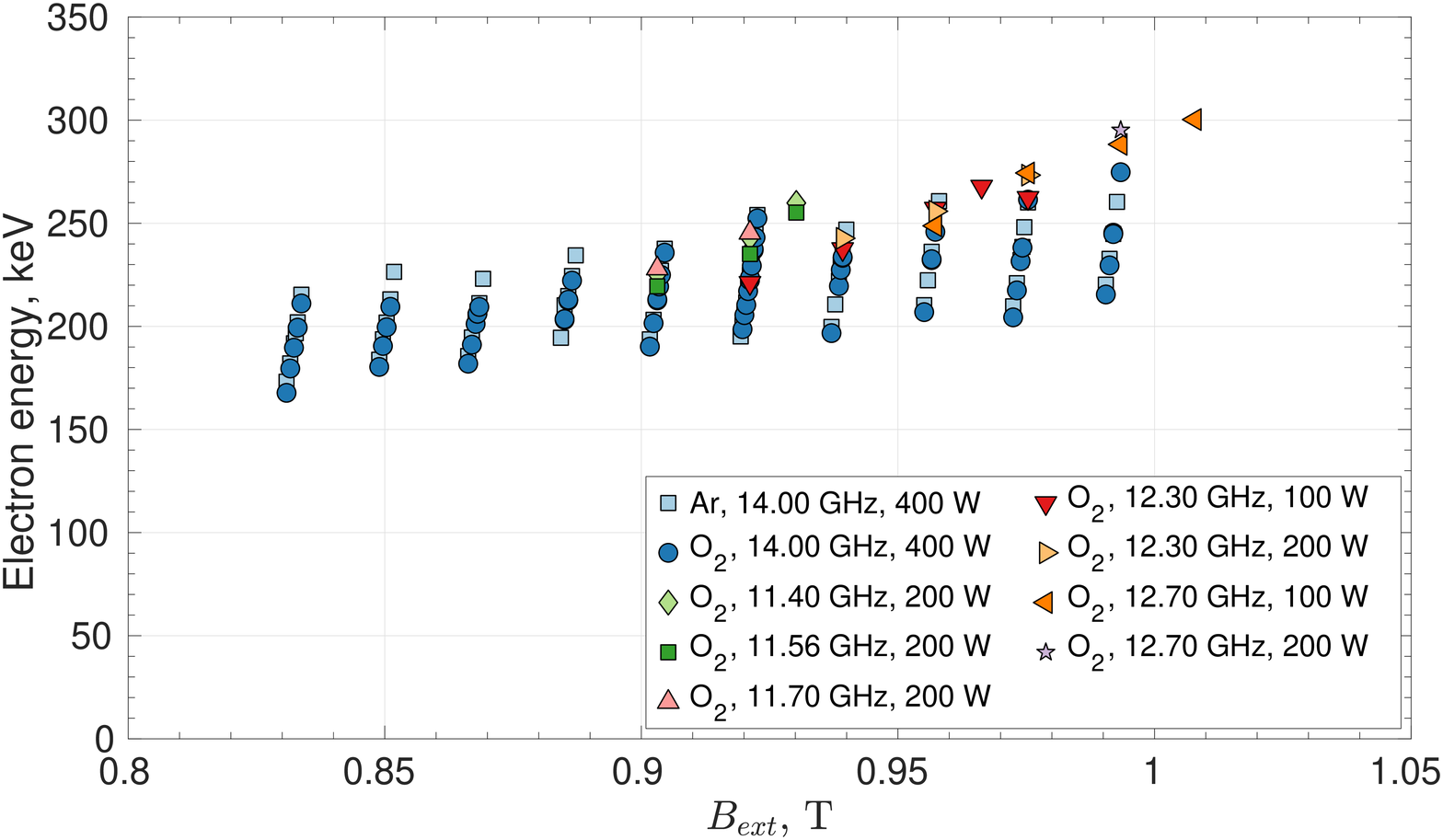}
\caption{\label{fig9}
The electron energy at the peak of the LEED hump as a function of $B_\mathrm{ext}$ at different settings.
}
\end{figure}

\begin{figure}
\includegraphics[width=\linewidth]{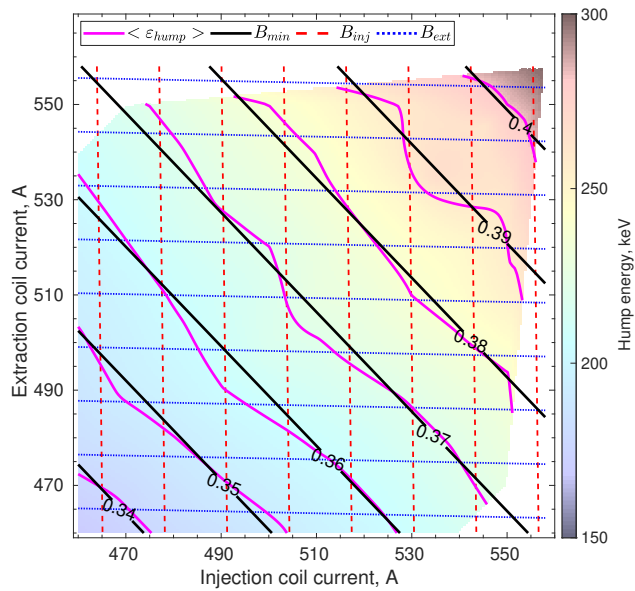}
\caption{\label{fig10}
The electron energy at the peak of the LEED hump (false color), its isolines (solid magenta), isolines of $B_\mathrm{min}$ (solid black), $B_\mathrm{inj}$ (dashed red), and $B_\mathrm{ext}$ (dotted blue) as a function of the current in injection and extraction coils. The data is the same as in Figs.~\ref{fig8} and \ref{fig9}.
Isolines of $B_\mathrm{min}$ are labeled.}
\end{figure}

It was found that only $B_\mathrm{min}$ had a clearly distinguishable effect on the hump energy, which increased with $B_\mathrm{min}$, as demonstrated in \fref{fig8}. As an example, the electron mean energy in the hump as a function of $B_\mathrm{ext}$ is shown in \fref{fig9}, in order to demonstrate its scatter. Finally, \fref{fig10} shows the energy of the hump (in false color, interpolated) against currents of the injection and extraction coils together with isolines of the hump mean energy (solid magenta), $B_\mathrm{min}$ (solid black), $B_\mathrm{inj}$ (dashed red), and $B_\mathrm{ext}$ (dotted blue). The data in \fref{fig10} underlines the fact that the energy of the hump follows the change of $B_\mathrm{min}$, not other magnetic-field-related quantities.

The observation suggests that there is a certain process that affects the formation of the hump on the distribution function of the lost electrons, that does not depend neither on the frequency or power of the heating radiation, nor on the gas type or the local mirror ratios (injection and extraction), but only on the absolute minimum value of the magnetic field in the center of the trap. Similar statement was made in \cite{LBNL-bremsstrahlung}, where the bremsstrahlung spectral temperature (being correlated, but not equal to the electron temperature) was found to depend on $B_\mathrm{min}$ and to be independent of the heating frequency (in a stable regime of operation, free of cyclotron instabilities). It should be explicitly pointed that the bremsstrahlung spectra in \cite{LBNL-bremsstrahlung} were those emitted from the confined plasma, whereas all data presented here relates to the escaping particles. However, similar dependencies suggest a strong correlation between the energy distributions of lost and confined electrons.

\subsection{Experiments with pulsed microwave injection}
\begin{figure}
\includegraphics[width=\linewidth]{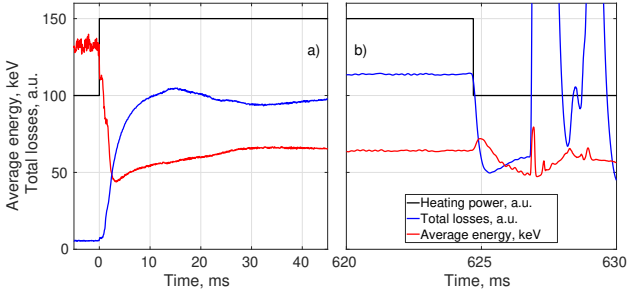}
\caption{\label{fig11}
Total losses and average energy evolution at the leading (a)  and the trailing (b) edge of the microwave pulse. Pulsed heating with 260 W at 14 GHz, 1 Hz repetition rate, duty factor 0.62. Oxygen pressure $3.5 \cdot 10^{-7}$ mbar, $B_\mathrm{min}/B_\mathrm{ECR}=0.735$.
}
\end{figure}

To determine the origin of the LEED hump visible at 170--300 keV the experiment with pulsed microwave injection was conducted with 14 GHz at 1 Hz pulse repetition rate and duty factor 0.6. The pulsed power was set to 260 W and the oxygen pressure to $3.5 \cdot 10^{-7}$ mbar. The coils were energized symmetrically with a current of 500 A, yielding $B_\mathrm{min}/B_\mathrm{ECR}=0.735$. The average energy of the lost electrons, electron flux and heating pulse (schematically) are shown as a function of time ($t=0$ corresponds to the leading edge of the microwave pulse) in \fref{fig11} for the beginning (a) and the end (b) of the heating pulse.

The electron flux rises abruptly (see \fref{fig11}a) immediately with the heating power from a very low yet non-zero level, experiencing a moderate overshoot and then slowly reaches the steady-state. The overshoot could be correlated with the preglow peak being thoroughly studied in \cite{PGW-1, PGW-2, PGW-3, PGW-PoP}. The non-zero electron flux before the microwave pulse can be explained by the heating power duty cycle. The time between consecutive pulses is short enough for electrons to remain confined in the trap for the whole time between pulses. At the trailing edge the flux drops abruptly together with the power (\fref{fig11}b). However several bursts are seen during the plasma decay, unambiguously matching with the afterglow kinetic instabilities \cite{Mansfeld-PPCF-2016,Afterglow-PoP-2012}, which are always present since regardless of the initial  plasma energy content defined by the ion source settings there will be a moment during the transient when the instability growth rate exceeds its damping rate \cite{Mansfeld-PPCF-2016, Shalashov-AFG,Shalashov-REV}. Following the instabilities the electron flux keeps decaying with a time constant of several tens of ms (not shown in the \fref{fig11}), which corresponds to the Coulomb scattering time for an energy of several keVs, being commonly considered as an average energy of electrons confined in ECR plasmas. The authors would like to point the Reader’s attention to the fact that the electron flux drops by a factor of two within less than one millisecond after the microwave power is switched off. This time is very short to be related to Coulomb scattering given the average energy of electrons, which suggests that the rf-induced scattering process \cite{RF-1,RF-2,SA-tokman} is comparable to  electron losses by collisional scattering in continuous operation mode of the ECR ion source.

Contrary to the electron flux, the average energy of the lost electrons drops with the leading edge of the microwave pulse, as the ECR heating starts to supply electrons with a wide energy spectrum. At the trailing edge of the microwave pulse the average energy exhibits some oscillations and then starts to rise slowly, finally reaching the value shown just before the consecutive pulse. The increase of the average energy is presumably related to the probability of Coulomb scattering decreasing with the electron energy, which implies that the confinement time of the high energy electrons is longer than the microwave off period in this case \cite{Olli-PSST-2009}.

\begin{figure}
\includegraphics[width=\linewidth]{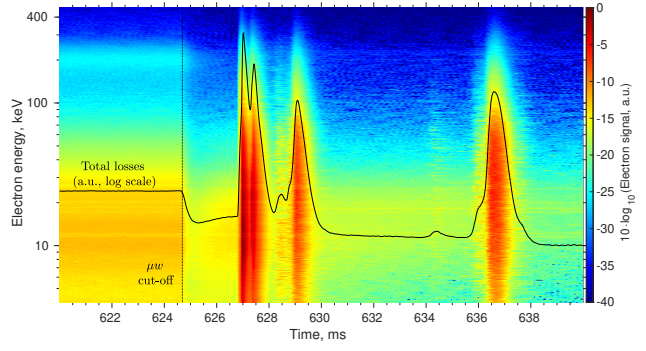}
\caption{\label{fig12}
The lost electrons energy spectrum as a function of time at the trailing edge of the microwave pulse. Total losses are shown (not to scale) with a black solid line. Parameters are the same as in \fref{fig11}.
}
\end{figure}

\begin{figure}
\includegraphics[width=\linewidth]{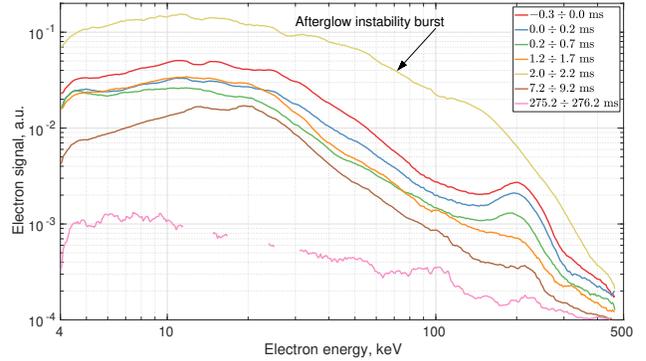}
\caption{\label{fig13}
The electron current (arbitrary units) as a function of the electron energy at the end of the heating pulse. Time origin is the trailing edge of the microwave pulse. Parameters are the same as in \fref{fig11}.
}
\end{figure}

The lost electrons energy spectrum as a function of time on both sides of the trailing edge of the microwave pulse is shown in \fref{fig12}. The electron flux is shown with false color. The afterglow instabilities are well visible, expelling significant number of electrons in a wide energy range at $t=627$, $627.5$, $629$ and $636.5$ ms. The moment of the microwave power being switched off is clearly seen in the spectrum and is also marked with a dotted vertical line. In addition to the drop of the total electron flux, discussed earlier, there is a noticeable change in the spectrum at energies above 100 keV related to the microwave power being switched off. The LEED hump is clearly visible at $\sim$200 keV when the microwaves are on, disappearing abruptly with the microwaves being turned off. This observation indicates that the microwave--electron interaction is responsible also for the existence of the hump in the LEED. Vertical slices of the spectrum at various times are shown in \fref{fig13}. The curves are plotted by integrating the signal in each energy bin within the time limits indicated in the plot legend (here $t=0$ ms corresponds to the microwave switch off). The data show that the hump at 200 keV disappears within less than 1.5 ms after the microwaves are switched off, which is not possible  to be explained by Coulomb scattering.

\begin{figure}
\includegraphics[width=\linewidth]{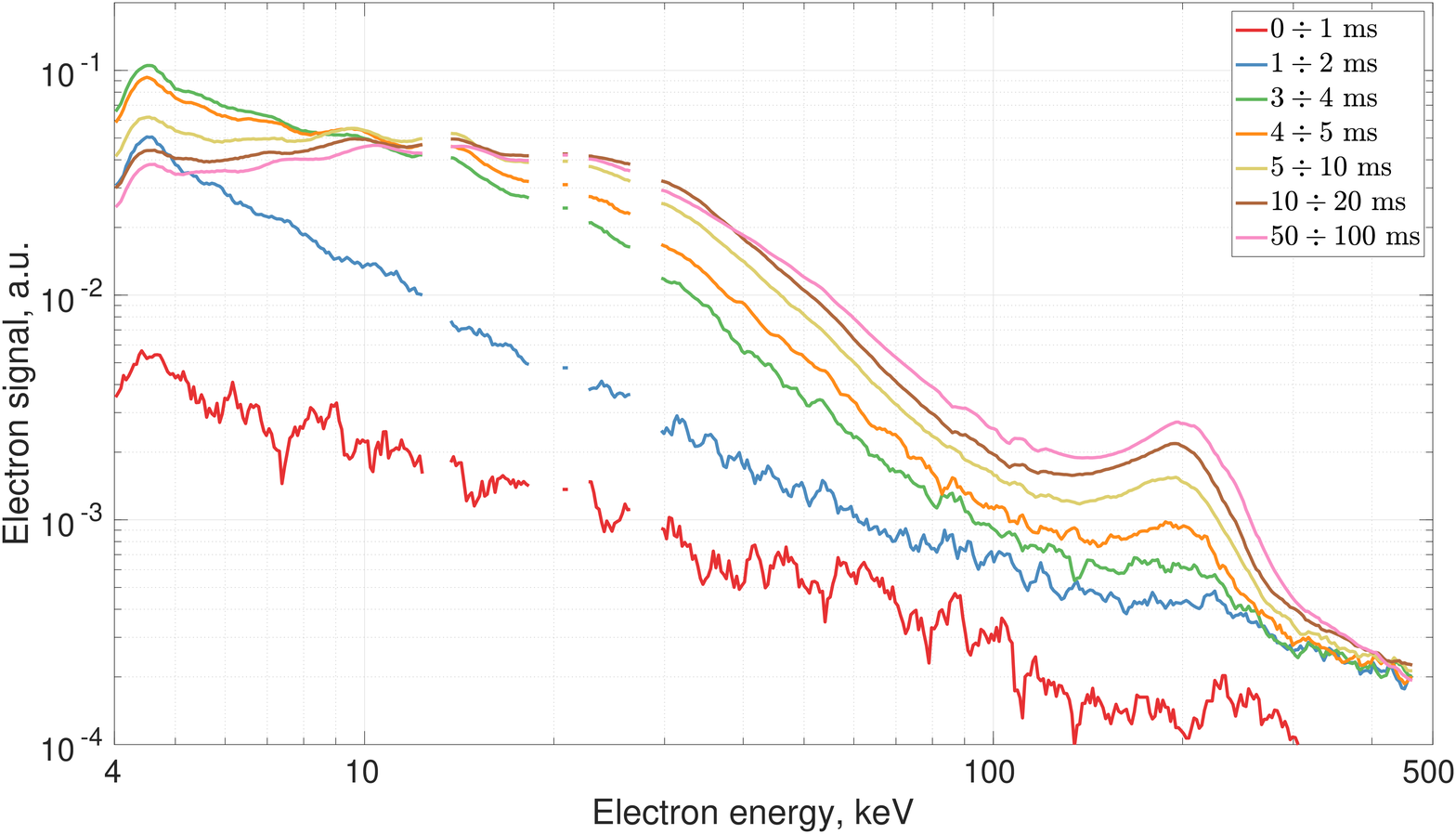}
\caption{\label{fig14}
The electron current (arbitrary units) as a function of the electron energy at the beginning of the heating pulse. Time origin is the leading edge of the microwave pulse. Parameters are the same as in \fref{fig11}.
}
\end{figure}

Similar slices of the energy spectrum but this time measured just after the leading edge of the microwave pulse are shown in \fref{fig14} ($t=0$ corresponds to the microwave switch on). Some data in the range 10--30 keV has been removed due to electrical noise. The curves in \fref{fig14} show the evolution of LEED, reaching the steady-state after $\sim100$ ms of microwave heating. The same characteristic time to reach a steady-state level is found for parameters such as bremsstrahlung and ion currents \cite{XRAY-2, PGW-3}. Although the hump at 200 keV appears almost immediately, it takes several tens of ms for it to reach the full magnitude. This indicates that the accumulation of electrons forming the discussed hump takes considerable time, which is consistent with the hypothesis of microwave heating being a prerequisite for the process populating the loss cone.

\section{Discussion}

\subsection{Electron trajectory tracing}

In an attempt to reproduce the measured energy distributions of the lost electrons, in particular the microwave-induced high-energy hump, we used the NAM-ECRIS code (Numerical Advanced Model of Electron Cyclotron Resonance Ion Source)  \cite{M1,M2}. 
The code simulates both the ion and electron dynamics in the ion source by running two separate modules that exchange the relevant information between each other.
The electron module traces the electron movement in the external magnetic field taking into account the electron-electron and electron-ion Coulomb collisions, the energy losses due to inelastic collisions with neutrals and ions (exitation and ionzation) and spontaneous emission of the electron-cyclotron radiation, the  heating by external microwave field. Inelastic collisions of electrons are modeled using excitation cross-sections from \cite{M5} and ionization cross-sections from \cite{M6}. To model the longitudinal confinement, electrons are reflected from the walls if their energy along the magnetic field lines is less than 50 eV, which is an order-of-magnitude estimate of the plasma potential.

The microwave heating is treated as ``kicks'' in electron velocity whenever it crosses the point where the ECR condition is satisfied.
The relativistic cyclotron resonance is
\begin{equation}\label{eq1}
s \omega_B^\mathrm{res}/ \omega = \gamma \left( 1 \pm \upsilon_\parallel/{\upsilon_\phi(\omega_B^\mathrm{res})} \right)
\end{equation}
where $s$ is the cyclotron harmonic number (fundamental or second), $\omega_B=eB/m_ec$ is the  cyclotron frequency of ``cold'' electron, $\omega$ is the heating wave frequency, $\gamma$ and $\upsilon_\parallel$ are the relativistic factor and  longitudinal  velocity of resonant electron, $\upsilon_\phi(\omega_B^\mathrm{res})$ is the wave phase velocity along the magnetic field line calculated at the resonance position, the positive and negative signs stand for the blue and red shifted resonances, correspondingly.
The wave phase velocity is calculated from the dispersion relation for the right-hand polarized (whistler) waves propagating presumably along the magnetic field in cold plasma as
\begin{equation}\label{eq2}
{c^2}/{\upsilon^2_\phi} \equiv n^2_\parallel = 1-{\omega^2_\mathrm{p}}/\left(\omega(\omega-\omega_B)\right),
\end{equation}
where $n_\parallel$ is the refractive index, $\omega_\mathrm{p}$ is the electron Langmuir (plasma) frequency. Then, the resonant surface is calculated following the spatial variation of the magnetic field strength $\omega_B(\mathbf{r})=\omega_B^\mathrm{res}$ or, equivalently, $B(\mathbf{r})=B_0\: \omega_B^\mathrm{res}/\omega$ with $B_0=0.5$ T for 14 GHz microwave heating.
The velocity ``kicks'' are calculated both perpendicular and along the local magnetic field line. The kick phase is a random value, whereas the kick magnitude is proportional to the local electric field. The simulation does not take into account spatial variations of the electric field, but rather treats it as a free parameter.

\begin{figure}
\includegraphics[width=\linewidth]{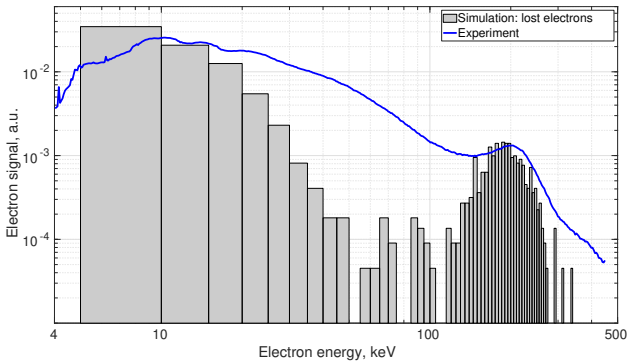}
\caption{\label{fig15}
Experimentally obtained LEED (solid blue line) and NAM-ECRIS code produced LEED (gray bars) for
argon plasma, argon pressure $3.5 \cdot 10^{-7}$, heated with 400 W at 14 GHz, $B_\mathrm{min}/B_\mathrm{ECR}=0.72$. LEEDs are normalized to unity square.
}
\end{figure}

Figure~\ref{fig15} shows the comparison of the experimental LEED and the corresponding distribution function simulated with NAM-ECRIS code. Distributions are normalized to the unity square. The model qualitatively reproduces the hump and its energy (close to 200 keV), whereas the overall shape of the LEED is not well described by the simulation. The origin of the hump modeled with NAM-ECRIS lies in the Doppler-shifted relativistic fundamental harmonic ECR, which apparently occurs close to extraction magnetic mirror for electrons with energies 200--300 keV heated by 14 GHz radiation, thus enhancing losses of such electrons. According to this approach, the hump energy should depend on the magnetic field near the extraction as well as the microwave frequency. However, the data in figures \ref{fig11}, \ref{fig12} and \ref{fig13} prove that the energy of the hump is independent of the extraction field and rather depends on the $B_\mathrm{min}$ value as described above. This fact implies that there is an additional process which is not taken into account by using the described approximation with the fundamental harmonic ECR heating.

It is worth noting that for relatively large plasma density the resonance condition for the second ECR harmonic is fulfilled close to $B_\mathrm{min}$ position: ${\frac{1}{2} B_0 \gamma \approx 0.36}$~T, yielding $\gamma=1.44$ and the energy of $\varepsilon\approx225$ keV, which is very close to the experimentally observed peak. However, if the peak would be caused by the second harmonic of the ECR heating, its energy would depend on both $B_\mathrm{min}$ and the heating frequency, i.e. on the value of $B_\mathrm{min}/B_\mathrm{ECR}$, which contradicts the experimental observations indicating that the hump position is not shifted with the frequency change if $B_\mathrm{min}$ remains constant.

\subsection{Quasilinear diffusion and rf-scattering}

In order to seek for an alternative mechanism possibly explaining the observed LEED hump we discuss the evolution of the electron distribution function in the framework of the quasilinear theory \cite{Tokman2,Tokman23,Tokman24}. Our application of the quasilinear theory to the open magnetic trap follows the reviewed methodology described in \cite{bible,bible1}. 

\begin{figure*}
\centering
\includegraphics[width=0.8\linewidth]{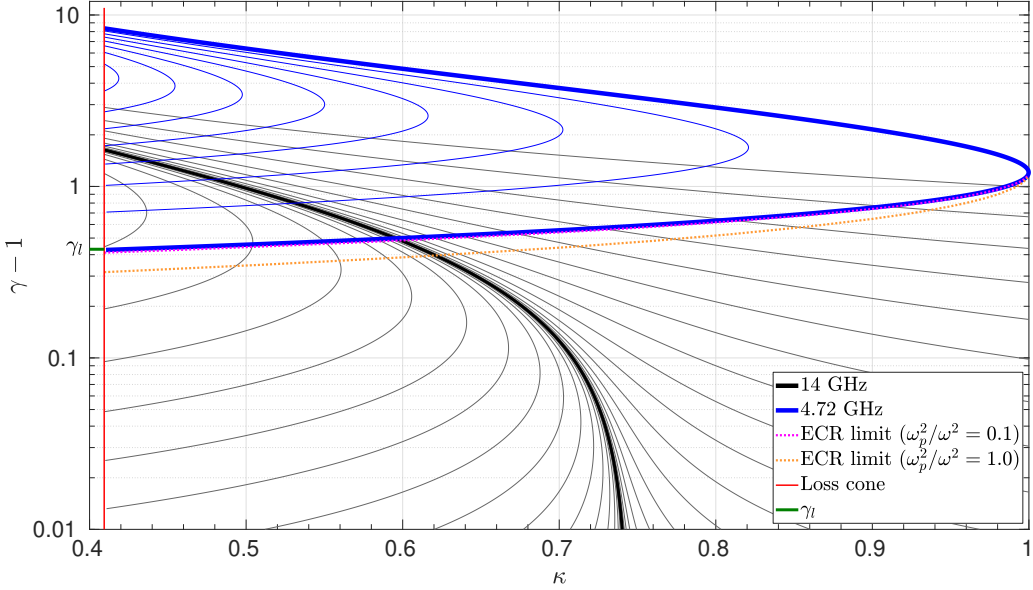}
\caption{\label{fig16}
The group of quasi-linear diffusion lines at 14 GHz (black) and 4.72 GHz (blue), loss cone (red), and ECR boundary for 4.72 GHz and $\omega^2_\mathrm{p}/\omega^2=0.1$ (solid orange) or $\omega^2_\mathrm{p}/\omega^2=1.0$ (dotted orange). Magnetic configuration: $B_\mathrm{min}=0.372$ T, $B_\mathrm{max}=B_\mathrm{ext}=0.908$ T. The minimum and maximum cyclotron frequencies are, accordingly, 10.4\;GHz and 25.4\;GHz. }
\end{figure*}

\begin{figure}
\includegraphics[width=\linewidth]{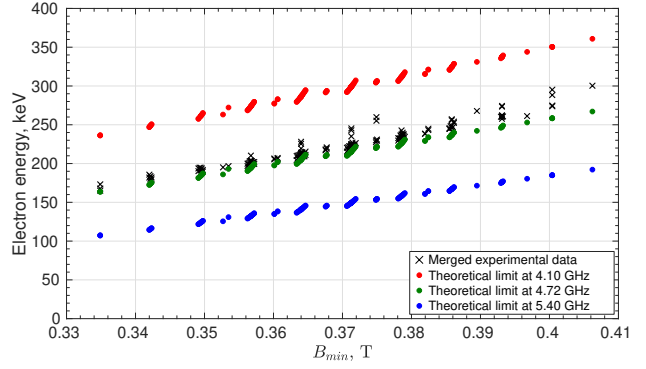}
\caption{\label{fig17}
The LEED hump energy from experiments (black crosses) and the corresponding limits of rf-induced losses calculated for 4.10 GHz (red dots), 4.72 GHz (green dots) and 5.30 GHz (blue dots) secondary frequency as a function of $B_\mathrm{min}$. The experimental data is taken from \fref{fig8}.
}
\end{figure}

Quasilinear approximation is plausible when the time of electron bounce oscillations is much longer than the time of a single pass of the ECR area, at which the change in the electron energy may be considered insignificant. For the ECR ion source this assumption is generally valid \cite{RF-3}. Then, it can be shown that the following quantity is conserved during interaction of a monochromatic wave with a collisionless electron: 
\begin{equation} \label{eq3}
\mathcal{K}=\varepsilon-\omega J_\perp=\mathrm{const},
\end{equation}
where $\varepsilon=(\gamma-1)m_e c^2 $ is the electron kinetic energy, $J_\perp = \gamma^2 \upsilon^2_\perp / 2 \omega_B(z)$ is the transverse adiabatic invariant. Here we assume that confined electrons are drifting (bouncing) along magnetic field lines, its longitudinal and perpendicular velocities vary such that $\varepsilon$ and $J_\perp$ remain constant; $\omega_B(z)$ denotes the gyrofrequency variation along a magnetic field line. Lines of $\mathcal{K} = \mathrm{const}$ are those  of quasilinear diffusion in the momentum space. The cyclotron interaction is reduced to the alignment of the electron distribution function along these lines, an example is shown in \fref{fig17} (explained hereafter). Such a process is usually limited either by the particle entering the loss cone or by reaching the energy and  adiabatic invariant values at which the ECR condition is no longer satisfied anywhere in the inhomogeneous magnetic field. Another possible mechanism for limiting the quasilinear diffusion is the so-called super-adiabatic effects \cite{SA-tokman}, which we do not consider below due to the steady-state plasma density in ECR ion source believed to be too high for superadiabatic effects.

Equation (\ref{eq3}) allows to determine the range of magnetic field strength, and therefore the range of $z$, allowed for a particle with given $(J_\perp,\gamma)$:
\begin{equation}\label{eq35}
\omega_B^\mathrm{min}\le\omega_B(z)\le\omega_B^*(J_\perp,\gamma)\equiv{ (\gamma^2-1)\: m_e c^2 }/2 J_\perp  .
\end{equation}
Here  $\omega_B^\mathrm{min}$ is the minimum gyrofrequency achieved at the trap center, $\omega_B^\mathrm{max}$ is the gyrofrequency at the electron mirroring point where $\upsilon_\parallel=0$. Deriving this equation from (\ref{eq3}) we assume $\mathcal{K}=0$ considering electrons accelerating from almost zero energies. Then, the ECR condition (\ref{eq1}) at the fundamental harmonic may be rewritten as
\begin{equation}\label{eq4}
{\omega_B^\mathrm{res}}/{\omega}-\gamma= \pm n_\parallel\sqrt{\left(\gamma^2-1\right)
\left(1-{\omega_B^\mathrm{res}}/{\omega_B^*}\right) } .
\end{equation}
Due to the strong slowing-down of the right-hand polarized waves in the vicinity of the cold resonance, i.e.\ $n_\parallel\gg1$ for $\omega_B\approx\omega$, condition \eref{eq4} may be fulfilled in a varying magnetic field for all electrons that could reach the cold ECR $\omega_B=\omega$ during their movement along the magnetic field line (bounce-oscillations), i.e.
\begin{equation}\label{eq5}
\omega_B^\mathrm{min}<\omega\le\omega_B^*(J_\perp,\gamma).
\end{equation}
This condition is valid when:
\begin{itemize}
\item  the plasma is not too rarefied, $\omega_\mathrm{p}^2/\omega_B^2\gtrsim \sqrt{T_e/m_ec^2}$ with $T_e$ being the characteristic temperature of the bulk electrons, so whistler-like dispersion relation \eref{eq2} works (this condition is valid in the reported experiments);
\item  the heating frequency is above the minimum gyrofrequency in the trap;
\item  the heating radiation occupies the whole plasma volume, rather than focused into a spot.
\end{itemize}
Then,  condition \eref{eq5} ensures that the electron with given $(J_\perp, \gamma)$ will meet the cyclotron resonance somewhere along the magnetic field line.
A similar condition may be written for the electron to enter the loss cone characterized with the maximum magnetic field value along a field line:
\begin{equation}\label{eq6}
\omega_B^\mathrm{max}\le\omega_B^*(J_\perp,\gamma),
\end{equation}
i.e.\ the turning point corresponds to non-existent magnetic field value, so the particle is able to leave freely the trap (the ambipolar potential is much less than kinetic energies considered here).

It is convenient to introduce a new  variable
\begin{equation}\label{eq_kappa}
\kappa\equiv \frac{\omega_B^\mathrm{min}}{\omega_B^*} \quad\left[\kappa=\frac{\omega_B^\mathrm{min}}{\omega_B}\frac{\upsilon^2_\perp}{\upsilon_\perp^2+\upsilon_\parallel^2}\right].
\end{equation}
Then, $\kappa=0$ and 1 correspond to electrons with no cyclotron gyration (freely moving along $B$-field) and no longitudinal motion (standing at the $B$-field minimum), accordingly. Conditions \eref{eq5} and \eref{eq6} becomes simply  vertical stripes in $(\kappa,\gamma)$-space: $\kappa\le \omega_B^\mathrm{min} / \omega_B^\mathrm{max}$ for the loss cone and $\kappa\le\omega_B^\mathrm{min} / \omega<1$ for the ECR presence. Summarizing, we state that in a dense enough plasma effective heating and confinement are possible for
\begin{equation}\label{eqf}
\omega_B^\mathrm{min} / \omega_B^\mathrm{max}\le\kappa\le {\omega_B^\mathrm{min}}/{\omega}<1,
\end{equation}
and this condition is independent of the electron kinetic energy $\gamma$.

Now we are ready to explain \fref{fig16} plotted in $(\kappa,\gamma)$-space. In solid black lines we show the group of diffusion lines  for the heating frequency of 14 GHz and $B_\mathrm{min}=0.372$ T, $B_\mathrm{max}=B_\mathrm{ext}=0.908$ T. The range of $\kappa$ defined by \eref{eqf} for these parameters is $0.41<\kappa\le 0.74$. The diffusion curves differ only by $\mathcal{K}$ value in \eref{eq3} that describes different initial conditions (velocity spread) for the accelerating electrons.  A quasi-one-dimensional distribution function localized along the line ${\mathcal{K}=0}$ (depicted with a bold black line) is formed from initially low-energy electrons. This diffusion asymptote lies entirely in the region fulfilling the inequality (\ref{eq5}) and crosses the loss cone when ${\gamma = 2\: \omega_B^\mathrm{max}} / \omega-1 = 2.44$ or $\varepsilon=736$ keV. The latter energy is much higher then the hump energy found in the experiment. Thus, the hump observed in every LEED shown above may not be solely explained by quasilinear diffusion as a result of interaction with the heating wave only. The arguments against the second ECR harmonics were discussed above. Moreover, the diffusion to the loss-cone along lines ${\mathcal{K}=\mathrm{const}}$ only would yield a delta function-like energy distribution of lost electrons, which is not the experimental observation.

In order to explain experimental data we must elaborate to model. Let us assume there exists a secondary electromagnetic wave with a frequency below the minimum cyclotron frequency. In \fref{fig16}, a group of diffusion lines for such a frequency (4.72 GHz) is plotted with blue color. The magnetic field is the same as for the 14 GHz diffusion lines (black curves). The non-resonant wave with ${\omega<\omega_B^\mathrm{min}}$ cannot be slowed down by the plasma significantly, see \eref{eq2}. In this particular case the argument of dense plasma is not applicable, so the resonance condition \eref{eq4} must be treated completely different. Indeed, for a limitied $n_{||}$ there is a minimum electron energy at which the interaction with a monocromatic wave is possible under the Doppler-shifted relativistic cyclotron resonance. One can qualitatively see that considering the low-energy limit, $\gamma\to1$, in \eref{eq4}: the left-hand-side stands positive and finite when ${\omega<\omega_B^\mathrm{min}}$, while the right-had-side goes to zero. After some algebra, one finds that condition \eref{eq4} may be fulfilled only if
\begin{eqnarray}\label{eq7}
\gamma\ge\gamma_\mathrm{min}(\kappa)=\frac{\omega_B^\mathrm{min} }{ \omega}\times\\
\nonumber\qquad\frac{1-n_\parallel\sqrt{(1-\kappa)\big(1-(1-n_\parallel^2(1-\kappa))\:(\omega/\omega_B^{\mathrm{min}})^2\big)}}{1-n_\parallel^2(1-\kappa)}.
\end{eqnarray}
In this equation, $n_\parallel$ must be calculated at the magnetic field minimum. It is important to note that the lower boundary of the electron energy depends weakly on the plasma density when $(\omega_\mathrm{p}/\omega)^2(\upsilon_\parallel/c)^2\ll 1$; when  $\omega_\mathrm{p} \to 0$ the energy boundary coincides with the particular diffusion line that is tangent to $\kappa=1$. This is illustrated in \fref{fig16}, the energy bottom boundaries are shown for ${\omega^2_\mathrm{p}/\omega^2=0.1}$ and 1.0 corresponding to the electron densities of $2.43 \cdot 10^{11}\;$cm$^{-3}$ and  $2.43 \cdot 10^{12}\;$cm$^{-3}$ (the cut-off density). The energy limit shifts upwards with lower plasma density, approaching the tangent diffusion line (bold blue line). This means that for diffusion lines shown in blue, the ECR condition is always met regardless of plasma density.

When the secondary electromagnetic wave is excited, the quasi-one-dimensional distribution function spreads along the blue diffusion lines corresponding to the additional frequency. In this case, two mechanisms of electron losses exist. The first one is the direct diffusion along blue lines when interacting with the second frequency. Secondly, diffusion along blue lines may ``transfer'' particles onto black diffusion lines corresponding to the main heating frequency, and these diffusion lines may then lead to the loss cone. Thus, adding the secondary frequency introduces additional losses of energetic electrons. Depending on the frequency $\omega_\mathrm{II}$ of the secondary wave, the additional electron losses may be introduced either in the whole energy range when $\omega_\mathrm{II}>\omega_B^\mathrm{min}$, or, when $\omega_\mathrm{II}<\omega_B^\mathrm{min}$, be more pronounced only at energies higher than some critical value $\gamma_{l}$ defined as a point where the diffusion line, which is tangent to $\kappa=1$, enters the loss cone $\kappa=\omega_B^\mathrm{min} / \omega_B^\mathrm{max}$. Analytical expression for $\gamma_{l}$ in plasma with a finite  density is quite bulky;  however, as this diffusion line is close to the lower energy limit for the ECR condition, one can approximate it with \eref{eq7} as
\begin{equation}\label{eq8}
	\gamma_{l}\approx \gamma_\mathrm{min}(\omega_B^\mathrm{min} / \omega_B^\mathrm{max}).
\end{equation}
This estimate for the characteristic energy of lost electrons posses the main feature of the experimental energy of the hump: it is proportional to $B_\mathrm{min}$ and weakly depends on all other parameters.

\begin{figure}
\includegraphics[width=\linewidth]{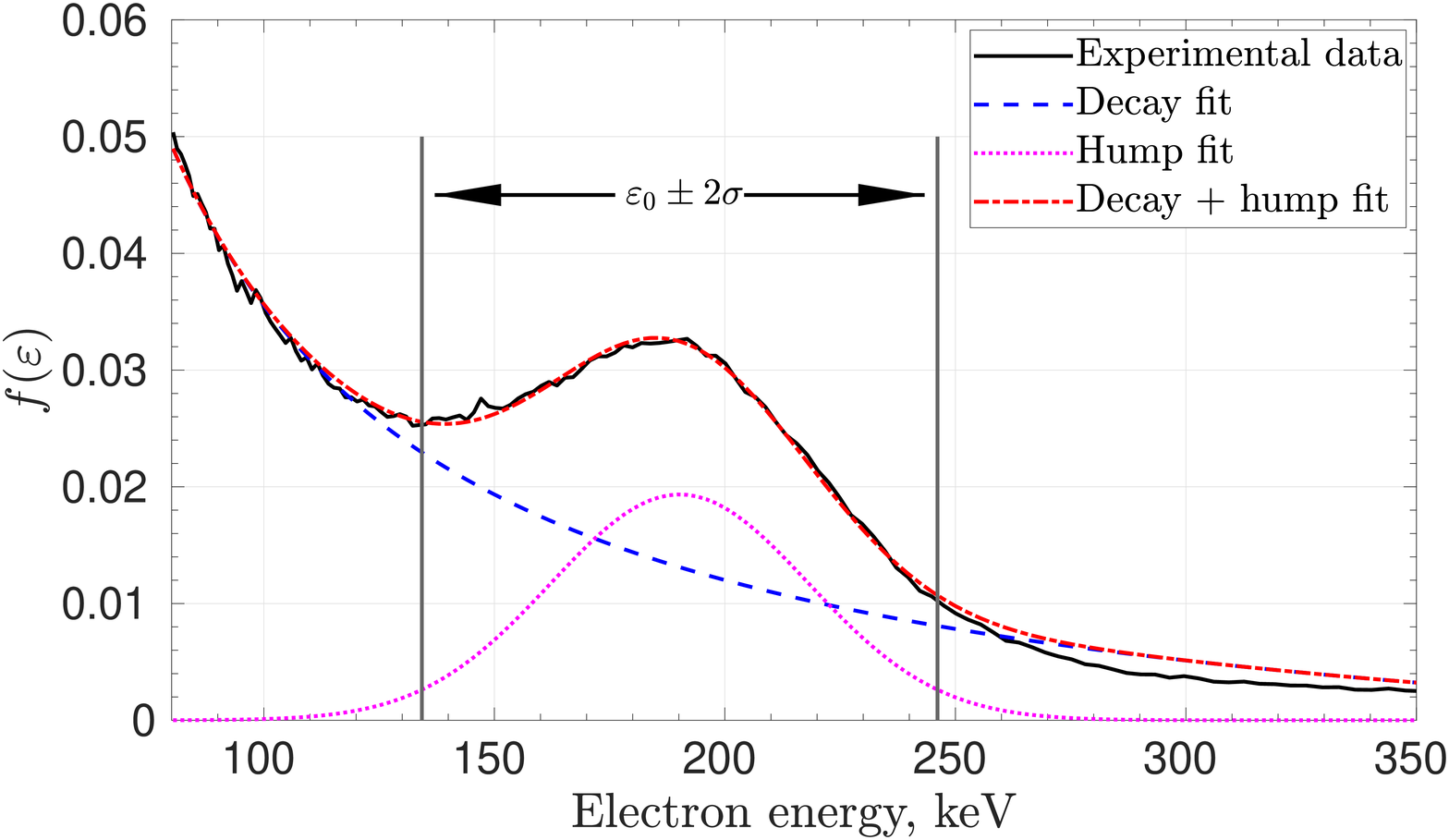}
\caption{\label{fig18}
The LEED hump approximation with Gaussian fit.
}
\end{figure}

Indeed, the experimentally observed dependence of the hump energy on $B_\mathrm{min}$  shown in \fref{fig8} is reasonably well reproduced in the frame of the above hypothesis assuming that the secondary frequency is constant, i.e.\ independent on any of ECR ion source parameters, and in the range of 4.10--5.40~GHz. Such comparison is shown in \fref{fig17}, where all the data from \fref{fig8} is merged and plotted with black crosses representing the electron energy at the peak of the hump. For each experimental data point, theoretical energy $ (\gamma_{l}-1)m_e c^2$  is estimated with the procedure explained above using three frequencies $\omega_\mathrm{II}$. The central frequency $\omega_\mathrm{II}/2\pi=4.72$ GHz was chosen because it is the closest eigenmode of the plasma chamber (TE$_{021}$), and the given range of frequencies covers 95\% of the electrons ($\pm 2\sigma$) in a Gaussian fit representing the hump after subtracting the exponentially decreasing background distribution function. The procedure for extracting the energy range of the hump is clarified in \fref{fig18} showing an example of a measured LEED and the hump after the background subtraction.

The described approach may explain the observed shape of LEED and its dependencies on parameters, though the origin of the secondary electromagnetic wave has to be yet defined. According to the experimental data, this supposed wave is turned on and off together with the heating radiation, though its frequency is independent of the primary heating frequency. One of possible candidates for such a wave might be the plasma-filled cavity mode excited by a strongly heated unstable plasma inside. It is then possible to explain the slow appearance and fast disappearance of the hump seen in figures~\ref{fig13} and \ref{fig14}. When the heating wave is turned on, it acts as a source of energetic electrons, so it takes some time to accumulate enough electrons above the instability threshold to build up the intensity of a broadband electromagnetic emission, exciting, probably spatially selected, a cavity mode near $\omega_\mathrm{II}$-frequency, which in turn forces these electrons to the loss cone. Quasilinear model suggests that in a stationary regime the population of resonant electrons that drive the $\omega_\mathrm{II}$-mode is kept close to the instability threshold \cite{Shalashov-REV,Trh,Shalashov-EPL}. Once the heating wave is turned off, the steady source of high-energy electrons disappears, and relaxation of the energetic electrons is then defined by $\omega_\mathrm{II}$-mode-induced scattering to the loss cone, which is a fast quasilinear process. At that, the population of the resonant electrons drops  below the threshold value during a short time interval, the $\omega_\mathrm{II}$-mode vanishes, and so does the LEED hump. Then the decay of the energetic electrons continues on much slower time-scale governed, most likely, by the Coulomb collisions.

\section{Conclusion}

The existence of a hump in the energy distribution of electrons escaping the magnetic confinement of the ECR ion source axially carrying more than 30\% of the energy (for electrons with energies above 4 keV) was demonstrated. It was found that the energy of the hump depends only on $B_\mathrm{min}$ and hardly on other quantities describing the magnetic field profile. Neither does it depend on the heating power and frequency or the gas pressure and composition.

A recent paper by Isherwood et al.\ \cite{Isherwood} reports a similar study of the high energy tail of the LEED and an observation of a hump at several hundred keV. They observed that the mean energy of the hump is insensitive to ion source parameters other than the axial magnetic field and found a correlation between the spectral temperature of the plasma bremsstrahlung and the mean energy of the LEED hump. However, Isherwood et al.\ present uncalibrated data, which is a major difference to the present work. The calibrated data presented here allows discussing possible mechanisms explaining the formation of the hump. Nevertheless, the findings are mutually corroborative and, thus, highlight the fact that the hump is a fundamental feature of the energy distribution escaping axially from a minimum-B ECR plasma.

The reported experimental data and its comparison to the numerical modeling allows excluding some plausible mechanisms, such as direct interaction with the heating wave and its harmonics. Within the frame of quasi-linear diffusion concept the hypothesis explaining the hump formation is proposed. The hypothesis is based on the assumption of the cavity mode excitation by the hot plasma. The excited mode then scatters energetic electrons to the loss-cone, forming the hump.

The aforementioned hypothesis, although not excluding other possible mechanisms, motivates further experiments including careful investigation of a electromagnetic emission of the magnetically confined plasma of an ECR ion source in the frequency range below the minimum cyclotron frequency.
There are several observations of  ECR plasma emission at discrete frequencies sufficiently lower than the heating frequency in the similar conditions \cite{Mansfeld-PPCF-2016,Izotov-PSST-2015,Izotov-PoP-2017};  the detection of these frequencies was attributed to the onset of cyclotron instabilities.
As well, discrete line emission in the relevant frequency range has been detected in the regime when no obvious indications of plasma instability is present \cite{Shalashov_PRL_2018}.

\section*{Acknowledgments}
The work, including data processing and analysis, was supported by the Russian Science Foundation, project No~19-12-00377; the experiment at JYFL facility was aslo supported by the Academy of Finland, project No~315855.

\section*{References}
\bibliography{ms}

\end{document}